\documentclass[aps,pre,twocolumn,showpacs,superscriptaddress,nofootinbib]{revtex4-2}
\usepackage[normalem]{ulem}

\usepackage{graphicx}% Include figure files
\usepackage{epstopdf}
\usepackage{calrsfs}
\usepackage{dcolumn}% Align table columns on decimal point
\usepackage{bm}% bold math
\usepackage{enumerate}
\usepackage{amsmath}

\usepackage[normalem]{ulem}  % \sout{old text} for strikeout
\usepackage[colorlinks]{hyperref}

\usepackage{amssymb}
\usepackage[mathscr]{eucal}
\usepackage{lipsum} % for filler text
\usepackage{ulem}

\begin{document}

%\preprint{APS/123-QED}

\title{Tilings of a bounded region of the plane by maximal one-dimensional tiles}

\author{Eduardo J. Aguilar}
\affiliation{Instituto de Ci\^ encia e Tecnologia,
Universidade Federal de Alfenas,
Po\c cos de Caldas, MG 37715-400, Brazil}
\author{Valmir C. Barbosa}
\affiliation{Programa de P\'os-Gradua\c c\~ao em Ci\^encias Computacionais, IME,
Universidade do Estado do Rio de Janeiro,
Rio de Janeiro, RJ 20550-900, Brazil}
\author{Raul Donangelo}
\affiliation{Instituto de F\'\i sica,
Universidade Federal do Rio de Janeiro,
Rio de Janeiro, RJ 21941-909, Brazil}
\affiliation{Instituto de F\'\i sica, Facultad de Ingenier\'\i a,
Universidad de la Rep\'ublica,
Montevideo 11.300, Uruguay}
\author{Welles A. M. Morgado}
\affiliation{ Departamento de F\'\i sica,
Pontif\'\i cia Universidade Cat\'olica,
Rio de Janeiro, RJ 22452-970, Brazil}
\affiliation{National Institute of Science and Technology---Complex Systems,
Teresina, PI 64049-550, Brazil}
\author{Sergio R. Souza}
\email{Corresponding author: srsouza@if.ufrj.br}
\affiliation{Instituto de F\'\i sica,
Universidade Federal do Rio de Janeiro,
Rio de Janeiro, RJ 21941-909, Brazil}
\affiliation{Departamento de F\'\i sica, ICEx,
Universidade Federal Fluminense,
Volta Redonda, RJ 27213-145, Brazil}

\date{\today}% It is always \today, today,

\begin{abstract}
We study the tiling of a two-dimensional region of the plane by $K$-cell
one-dimensional tiles, or $K$-mers. Unlike previous studies, which typically
allowed for one single value of $K$ or sometimes a small assortment of fixed
values, here a tiling may concomitantly employ $K$-mers comprising any number
$K$ of cells, provided a maximality constraint is satisfied. In essence, this
constraint requires each of the $K$-mers in use to be as lengthy as possible,
given its surroundings in the resulting tiling. Maximality aims to limit the
variety of possible tilings while allowing for interesting behavior in terms of
the statistical physical observables of interest. In fact, by introducing an
energy function based on cell contacts and parameterizing it appropriately, we
have been able to observe relatively unexpected behavior, including the
suggestion of phase transitions as the system's temperature evolves. 

\end{abstract}

\maketitle

%%%%%%%%%%%%%%%%%%%%%%%%%%%%%%%%%%%%%%%%%%%%%%%%%%%%%%%%%%%%%%%%%%%%%%%%%%%%%%%%%%%%%%%

\begin{section}{Introduction}
\label{sect:introduction}

Given a finite collection $C$ of tile types, a tiling of the plane is a
covering that employs tiles from $C$ only and allows for no superposition of
tiles and no gaps between them. For a long time, probably beginning in the early
17th century with Kepler, then stretching through a gap of inactivity that
lasted through the 1970s and another through the 2020s, tilings constituted a
subject of almost exclusively mathematical interest. The main problem was to
find the smallest possible set $C$ that could tile the plane aperiodically
(i.e., with endless pattern diversification as the only possibility).
The first serious candidates are due to Penrose, first with six edge-marked tile
types that completed Kepler's early attempts \cite{p74}, then with two tile
types \cite{p78}. It took another half century for
Smith et al.~\cite{smkg24a,smkg24b} to reveal the first solutions with one
single tile type.

One of the two-tile type sets from Ref.~\cite{p78} has had a lasting impact in
the field of crystallography, where it suggested structural ordering outside the
classical approach \cite{m82}, prefiguring the discovery of quasicrystals
\cite{sbgc84} and influencing the field ever since \cite{zgbcult23}. Other
applications of the Penrose tile sets include, e.g., the study of graph-theoretic
properties of the classical dimer model \cite{fsp20}. Moreover, the very notion
of ``aperiodicity'' underlying them has been strongly influential, e.g., in
interpreting the results of self-assembled crystal structures from molecular
building blocks \cite{pd23}, and demonstrating the use of ``algorithmic''
self-assembly of DNA strands into cellular automata.\footnote{Specifically, one
based on Wolfram's elementary rule 90, the XOR rule \cite{w83}; see
Ref.~\cite{rpw04} and references therein.} Self-assembled systems are now part
of cutting-edge research in fields like materials science \cite{klclhjh22} and
DNA-based computing \cite{wdmhzyw19,dfglllnssyz21,xcs22,kljlklkck23}.

Issues of periodicity, however, become meaningless when the focus shifts toward
tilings of bounded regions of the plane, which is the case with applications
like modeling biological tissues \cite{fraej07,ags17,bhws17} and their
properties \cite{csmbg20,pkbmqtpmkgnsbrkthsiwthwmbdf15}, and developing
bioinspired metamaterials \cite{p21}. Given finiteness, the central issue is to
count the number of distinct tilings that the tile-type collection $C$ admits.
Such counting opens the way to calculating all sorts of useful statistical
properties, though one must reckon with the fact that only very rarely is it
obtainable from a closed-form expression. In this work, we consider $K$-cell
one-dimensional tiles, or $K$-mers, exclusively.

Given a rectangular region of $m\times n$ square ($1\times 1$) cells,
already for arbitrary $m,n\ge 2$ it seems that a closed-form
expression for how many distinct tilings $C$ admits is known only
if $C$ comprises dimers exclusively \cite{k61,tf61}. Beyond this, fixing $m=2$
and leaving $n$ unconstrained while monomers and dimers are the allowed tile
types seems to be as far as one can go with recurrence relations and generating
functions (see, e.g., Ref.~\cite{ks09}). These limitations notwithstanding, two
powerful approaches for asymptotic analysis have been used successfully on
rectangular regions: the transfer-matrix method (assuming the region of interest
either to be a lengthy strip \cite{b92} for tiling with Penrose tiles, or to have
some form of periodic boundaries on one of the dimensions---see
Refs.~\cite{gd07} for trimers, \cite{rsd23} for $K$-mers with fixed $K$, and
references therein); and the Wang-Landau method \cite{wl01a,wl01b} for state
(tiling) density estimation (see Ref.~\cite{tiling2025} for assorted $K$-mers
with $K=1,2,4$).

Here we consider the tiling of the $m\times n$ region when $C$ contains every
$K$-mer for $1\le K\le \max\{m,n\}$. We constrain the use of any one tile by the
following maximality condition. Placing a tile horizontally requires it to be
abutted at each end by a vertically positioned tile; placing the tile vertically
requires the abutting tiles to be positioned horizontally. We also assume
periodic boundary conditions on both dimensions, which immediately precludes
the use of an $m$-mer vertically or an $n$-mer horizontally while at the same
time abiding by  maximality. Thus, given periodic boundaries, we revise the
set $C$ to contain every $K$-mer for $1\le K\le \max\{m,n\}-1$, along with two
circular bands, one with $m$ cells and one with $n$ cells. Taken together, the
maximality requirement and the assumption of periodic boundaries have the
potential of simplifying the state space, and consequently allowing a cleaner
application of the transfer-matrix method, which seems like a natural choice in
this case.

Given an assignment of Ising-like states to the $m\times n$ cells (state $-1$ or
$+1$), each resulting $K$-mer, with its own particular value of $K$, is a
sequence of $K$ cells with like states, arranged horizontally (all cells in
state $-1$) or vertically (all cells in state $+1$), provided tile
maximality is respected. Likewise, cells in each resulting $m$-cell band are all
in state $+1$ and cells in each resulting $n$-cell band are all in state $-1$.
In all our analyses, an energy function is used that
adds up contributions from every pair of juxtaposed cells. Each of the possible
contribution values depends on four model parameters, viz., $f_L$, $f_U$,
$g_L$, and $g_U$, taking into account the two cells' states ($f_L$ and $g_L$ for
like states, $f_U$ and $g_U$ for unlike states) and juxtaposition directions
($f_L$ and $f_U$ for horizontal, $g_L$ and $g_U$ for vertical). This allows for
a rich variety of systems to be represented and likewise for several observables
to be studied.  

We continue by detailing our theoretical framework in Section~\ref{sect:theory},
presenting our results in Section~\ref{sect:results}, and concluding in
Section~\ref{sect:conclusions}.

\end{section}

%%%%%%%%%%%%%%%%%%%%%%%%%%%%%%%%%%%%%%%%%%%%%%%%%%%%%%%%%%%%%%%%%%%%%%%%%%%%%%%%%%%%%%%
 
\begin{section}{Theoretical framework}
\label{sect:theory}
The tilings consist of two qualitatively different tiles: horizontal and vertical,  as illustrated in Fig.~\ref{fig:tiles}.
The former are associated with cell spins $s=-1$ whereas the latter with $s=+1$, in analogy with the Ising model.
The surface on which the tiles are arranged consists of $m$ rows and $n$ columns.
Since the transfer matrix method is adopted throughout this work, periodic boundary conditions are imposed along the horizontal axis.
For simplicity, the vertical axis is also wrapped around, so the tiling is performed on a toroidal geometry.

\begin{figure}[t] 
    \centering
    \includegraphics[width=0.3\textwidth]{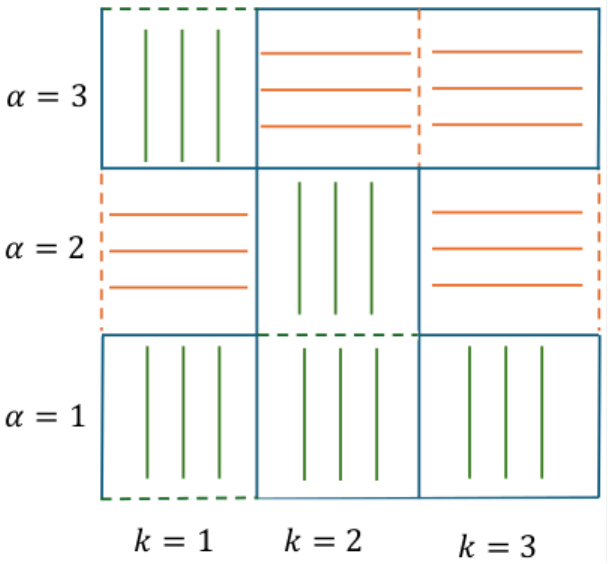}
    \caption{Example of a tiling for $m=n=3$. The coordinates are chosen so that the origin is located at the bottom left corner. Cell state $-1$ is represented
by horizontal stripes, cell state $+1$ by vertical stripes. The row and column indices increase from left to right and from bottom to top, respectively. Owing to the periodic boundary conditions, two horizontal and two vertical dimers are present in this configuration.
 For details, see the text.} 
    \label{fig:tiles} 
\end{figure}

%%%%%%%%%%%%%%%%%%%%%%%%%%%%%%%%%%%%%%%%%%%%%%%%%

\begin{subsection}{Energy}
\label{subsect:projstat}
%%%%%%%%%%%%%%%%%%%%%%%%%%%%%%%%%%%%%%%%%%%%%%%%%
The interaction between the different constituents of the system is restricted to nearest neighbors, i.e., a cell located at row $\alpha$ and column $k$, $(\alpha,k)$,
interacts only with those at $(\alpha \pm 1,k)$ and $(\alpha,k\pm 1)$.
For simplicity, there is no self-energy term.
By denoting the configuration of a row $\alpha$ as $\mu_\alpha=\left\{s_1^{(\alpha)},s_2^{(\alpha)},\cdots,s_n^{(\alpha)}\right\}$, the energy associated with
two adjacent rows is written as
\begin{equation}
E(\mu_{\alpha},\mu_{\alpha+1})=\frac{1}{2}\sum_{k=1}^n\left\{h_{k,k+1}^{(\alpha)}+h_{k,k+1}^{(\alpha+1)}\right\}+\sum_{k=1}^n v_{\alpha,\alpha+1}^{(k)}\;,
\label{eq:Erows}
\end{equation}
where
\begin{align}
h_{k,k+1}^{(\alpha)}=\frac{f_L}{4}&(1+s_{\alpha,k})(1+s_{\alpha,k+1})\nonumber\\
+\frac{f_U}{4}[&(1+s_{\alpha,k})(1-s_{\alpha,k+1})\nonumber\\
 +&(1-s_{\alpha,k})(1+s_{\alpha,k+1})]
\label{eq:h}
\end{align}
and
\begin{align}
v_{\alpha,\alpha+1}^{(k)}=\frac{g_L}{4}&(1-s_{\alpha,k})(1-s_{\alpha+1,k})\nonumber\\
+\frac{g_U}{4}[&(1+s_{\alpha,k})(1-s_{\alpha+1,k})\nonumber\\
 +&(1-s_{\alpha,k})(1+s_{\alpha+1,k})]\;.
\label{eq:v}
\end{align}
Above, $f_L$, $f_U$, $g_L$, and $g_U$ are model parameters.
They are intended to mimic the interaction between two adjacent pairs and the corresponding contribution to the system energy.
The subscripts $L$ and $U$ refer to interactions between two like and unlike cells, respectively.
The energy term $E(\mu_\alpha)$ corresponding to pairs lying along the row $\alpha$ is not explicitly included, in order to avoid
double counting.
In this way, the total energy of the system is obtained by summing the terms given by Eq.~(\ref{eq:Erows}) over all adjacent row pairs.

One should note that the cells $(1,2)$ and $(2,2)$ in Fig.~\ref{fig:tiles} form a maximal tile and give a null contribution to the total energy.
In contrast, the pair $(1,1)$-$(1,2)$ contributes $f_L$ to it.
In the same vein, the pair $(3,2)$-$(3,3)$ also forms a maximal tile, giving no contribution to Eq.~(\ref{eq:Erows}), whereas the pair $(2,3)$-$(3,3)$ adds $g_L$ to it.
Furthermore, cell pairs made up of different tiles along the vertical direction, such as $(1,3)$-$(2,3)$, add $g_U$ to the total energy, whereas those
along the horizontal direction, such as the pair $(2,2)$-$(3,2)$, contribute  $f_U$.
Thus, a convenient choice for the parameters enables one to study the properties of isotropic and anisotropic systems.
Finally, due to the periodic boundary conditions, pairs $(1,1)$-$(1,3)$ and $(2,1)$-$(2,3)$ form maximal tiles, while the pairs $(1,2)$-$(3,2)$ and $(1,3)$-$(3,3)$ contribute $g_U$ each to the total energy, and $(3,1)$-$(3,3)$ adds $f_U$ to it.

\end{subsection}

%%%%%%%%%%%%%%%%%%%%%%%%%%%%%%%%%%%%%%%%%%%%%%%%%

\begin{subsection}{Observables}
\label{sect:Observs}
%%%%%%%%%%%%%%%%%%%%%%%%%%%%%%%%%%%%%%%%%%%%%%%%%
The above formulation allows one to write the element $\langle \mu_\alpha | P | \mu_\alpha'\rangle$ of the transfer matrix $P$ as
\begin{equation}
\langle \mu_\alpha | P | \mu'_\alpha\rangle=e^{-\beta E(\mu_\alpha,\mu'_{\alpha+1})}\;,
\label{eq:tm}
\end{equation}
where $\beta=1/T$ and $T$ is the canonical temperature.
The Helmholtz free energy may thus be written as
\begin{equation}
F(T)=-m T \log(\lambda_1)\;,
\label{eq:F}
\end{equation} 
where $\lambda_1$ symbolizes the largest eigenvalue of the transfer matrix.
We take $m=50$ throughout this work, except where stated otherwise.
This value is sufficiently large to allow us to neglect the further terms associated with the smaller eigenvalues of the transfer matrix in the Helmholtz free energy without qualitatively affecting the quantities we are interested in.

Owing to the complexity of the transfer matrix, the eigenvalues cannot be obtained analytically, so we must use discrete values of $T$ within a convenient range.
This range depends on the parameters used in the calculation of the system's energy.
The range $0.1 \le T \le 10.0$ with steps of $\delta T=0.1$ is sufficient to highlight the relevant features of the underlying physics in most of the points addressed in this work.

\begin{figure}[t] 
    \centering
    \includegraphics[width=0.4825\textwidth]{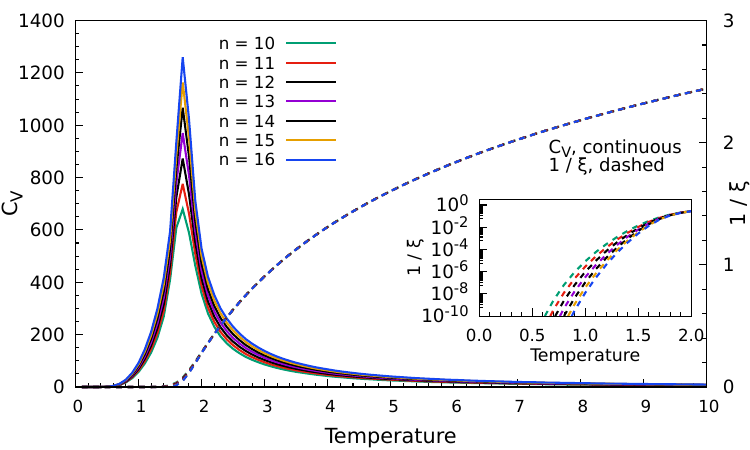}
    \caption{Left axis: Heat capacity for different system sizes, obtained with $f_L=1, f_U=2, g_L=1,$ and $g_U=2$. Right axis: The inverse of the correlation length for the same system sizes and parameters. Continuous lines represent the heat capacity while dashed lines depict the inverse of the correlation length.
 For details, see the text.} 
    \label{fig:cv12_12} 
\end{figure}

\begin{figure}[t] 
    \centering
    \includegraphics[width=0.4825\textwidth]{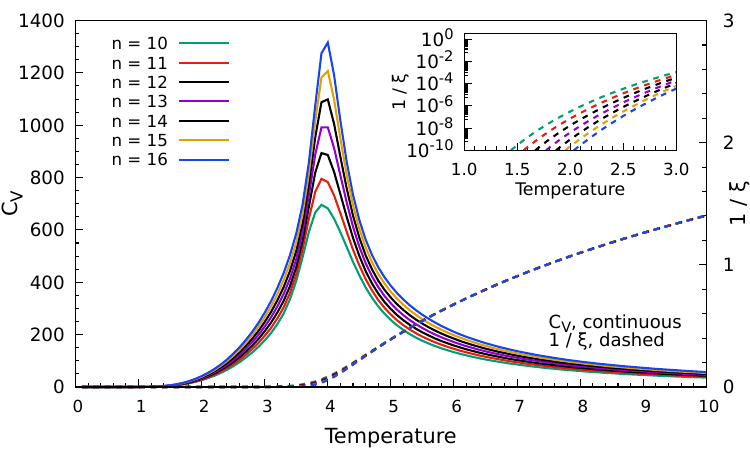}
    \caption{Same as Fig.~\ref{fig:cv12_12}, now for $f_L=1, f_U=4,g_L=1$, and $g_U=4$.} 
    \label{fig:cv14_14} 
\end{figure}

\begin{figure}[t] 
    \centering
    \includegraphics[width=0.4825\textwidth]{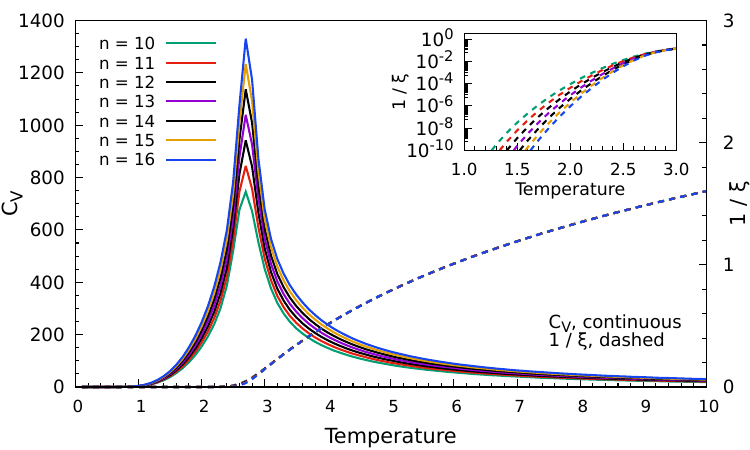}
    \caption{Same as Fig.~\ref{fig:cv12_12}, now for $f_L=1, f_U=2,g_L=1$, and $g_U=4$.} 
    \label{fig:cv12_14} 
\end{figure}

\begin{figure}[t] 
    \centering
    \includegraphics[width=0.4825\textwidth]{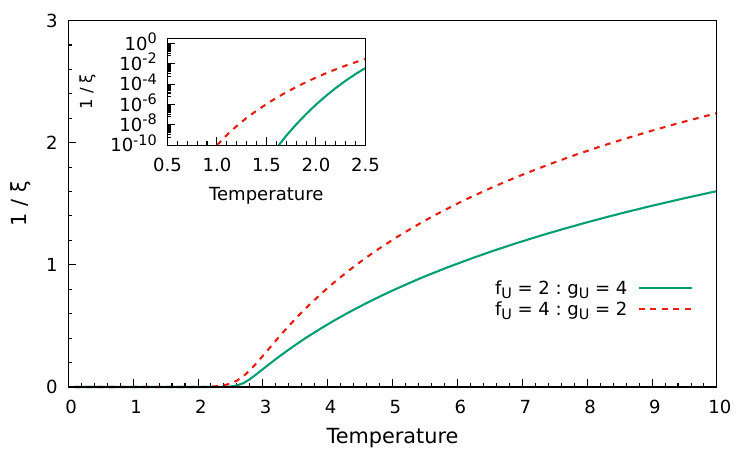}
    \caption{Inverse of the correlation length as a function of temperature for $n=m=16$ and different values of $f_U$ and $g_U$ (see legends).
In both cases, $f_L=g_L=1$. 
 For details, see the text.} 
    \label{fig:cor_exchange} 
\end{figure}

\label{sect:model}

\end{subsection}

\
 
\end{section}

\

\begin{section}{Results}
\label{sect:results}
%%%%%%%%%%%%%%%%%%%%%%%%%%%%%%
We begin by examining the heat capacity of the system, given by
\begin{equation}
C_V=-T\left(\frac{\partial^2 F}{\partial T^2}\right)_V\;,
\label{eq:cv}
\end{equation}
for $V=mn$, which is easily calculated with the help of Eq.~(\ref{eq:F}) and the standard numerical formulas for derivatives \cite{AbS72}.
Its behavior as a function of $T$ is displayed in Fig.~\ref{fig:cv12_12}, for different system sizes and the parameter set $f_L=g_L=1$, and $f_U=g_U=2$.
It reveals that $C_V$ is sharply peaked at $T_c\approx 1.7$, and that the height of the peak increases nearly linearly with system size.
These features suggest the occurrence of a second-order phase transition.
To examine this aspect more closely, Fig.~\ref{fig:cv12_12} also displays the inverse correlation length, $\xi^{-1}$.
This quantity is calculated from the eigenvalues of the transfer matrix as in
\begin{equation}
\xi=1/\log(\lambda_1/\lambda_2)\;,
\label{eq:xi}
\end{equation}
where $\lambda_2$ corresponds to the second largest eigenvalue. 
The results show that $\xi^{-1}$ is large at high temperatures, meaning the system is in disordered states.
However, the correlation length, $\xi$, rapidly increases as the temperature approaches $T_c$,  which is a typical signature of
a phase transition and indicates that the system is reaching ordered configurations.
Nevertheless, the linear scale makes it difficult to assert whether the divergence will also take place at $T=T_c$.
The inset shown in the figure seems to indicate that this might occur, as the divergence moves toward $T_c$ with increasing system size.
The differences are likely due to finite system-size effects.
We come back to this point below.

\end{section}

To examine the sensitivity of our findings to model parameters, we increase the cost of having a tiling with unlike neighbors by setting $f_U=g_U=4$, while keeping $f_L=g_L=1$.
The corresponding results, shown in Fig.~\ref{fig:cv14_14}, reveal that the same qualitative features persist.
The main differences are a shift in the critical temperature to $T_c\approx  4.0$ and broader peaks.
We find the same behavior for the correlation length as observed previously.
Therefore, these results also suggest that this system undergoes a phase transition as it cools down to $T_c$, the latter depending on the model parameters.

Among the different qualitative tilings, it is possible to imagine a surface where vertical tiles are more lenient toward horizontal neighbor tiles and vice versa.
The results for the parameter set $f_L=1, f_U=2,g_L=1$, and $g_U=4$ are shown in Fig.~\ref{fig:cv12_14}.
Once more, the same qualitative features are observed, which gives support to the robustness of our findings.
We do not show the symmetric case, $f_L=1, f_U=4,g_L=1$, and $g_U=2$, as it would lead to the same qualitative conclusions regarding the global thermodynamic observables.
This is due to the fact that the total energy depends on the total number of bonds, which is given by the sum over all vertical and horizontal interactions.
Owing to the periodic boundary conditions, the number of horizontal and vertical bonds is identical if the same number of rows and columns are employed.
By making $f_U\leftrightarrow g_U$, it is only the costs between bonds that are exchanged.
Since the order in the sum of the terms in the total energy is not relevant, the partition function is not affected.
We have checked that the quantitative differences are due to the doughnut geometry of the surface and that strictly the same global thermodynamic quantities are obtained, if we use the same number of rows and columns.

On the other hand, the correlation length has directional properties as it propagates the correlations from one row to another and the Hamiltonian of our model is not symmetric under a 90$^\circ$ rotation.
Therefore, one should expect differences in $\xi$ by reversing $f_U\leftrightarrow g_U$.
This is indeed observed in Fig.~\ref{fig:cor_exchange}, where we show the results for $n=m=16$, so that the differences must be attributed to the symmetry breaking inherent to the Hamiltonian.

%Among the different kinds of qualitative tilings, it is also possible to imagine a surface where vertical tiles are more lenient toward horizontal neighbor tiles and vice versa.
%The results for these two different situations are exhibited in Figs. \ref{fig:cv12_14} and \ref{fig:cv14_12}.
%In the former case, the parameters are set to $f_L=1, f_U=2,g_L=1$, and $g_U=4$, while in the latter case, they are $f_L=1, f_U=4,g_L=1$, and $g_U=2$.
%Once more, the same qualitative features are observed, revealing the they are intrinsic properties of the model and not fortuitous findings.

%\begin{figure}[h!] 
 %   \centering
 %  \includegraphics[width=0.4825\textwidth]{cv14_12.pdf}
%    \caption{Same as Fig.~\ref{fig:cv14_12} for $f_L=1, f_U=4,g_L=1$, and $g_U=2$. 
% For details, see the text.} 
%    \label{fig:cv14_12} 
%\end{figure}

\begin{figure}[t] 
    \centering
    \includegraphics[width=0.4825\textwidth]{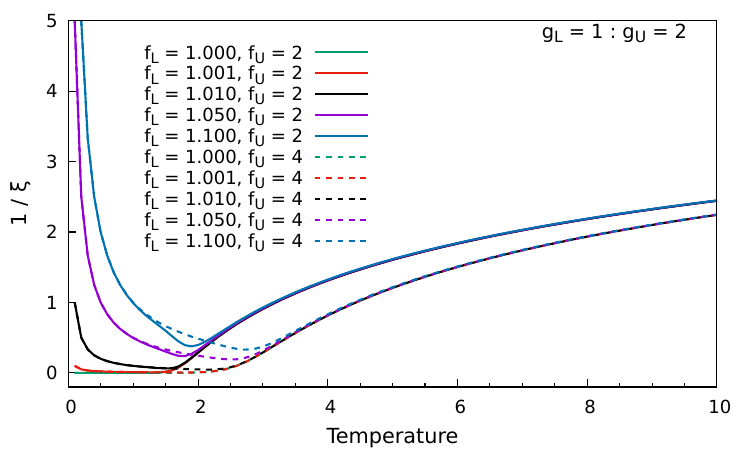}
    \caption{Inverse of the correlation length as a function of temperature for $n=10$ and different parameter values. 
 For details, see the text.} 
    \label{fig:xi_f} 
\end{figure}

A closer inspection of the correlation length provides further information about the characteristics of our system.
Figure \ref{fig:xi_f} shows $\xi^{-1}$ as a function of the temperature for slightly different values of $f_L \ge 1$, while the other parameters are kept fixed.
We first note that the behavior at high temperature is dominated by $f_U$.
More precisely, a higher cost for having unlike neighbors leads to a more ordered system, as mixed tiles would add a significant number of terms proportional to $f_U$ to the total energy.
At lower temperatures, the curves are grouped according to the cost of having similar neighbors.
This is consistent with a phase transition where ordered configurations tend to dominate.
Nevertheless, the system's tendency toward an ordered phase vanishes as $T \rightarrow 0$, with the exception of the case where $f_L=1$. 

\begin{figure}[t] 
    \centering
    \includegraphics[width=0.4825\textwidth]{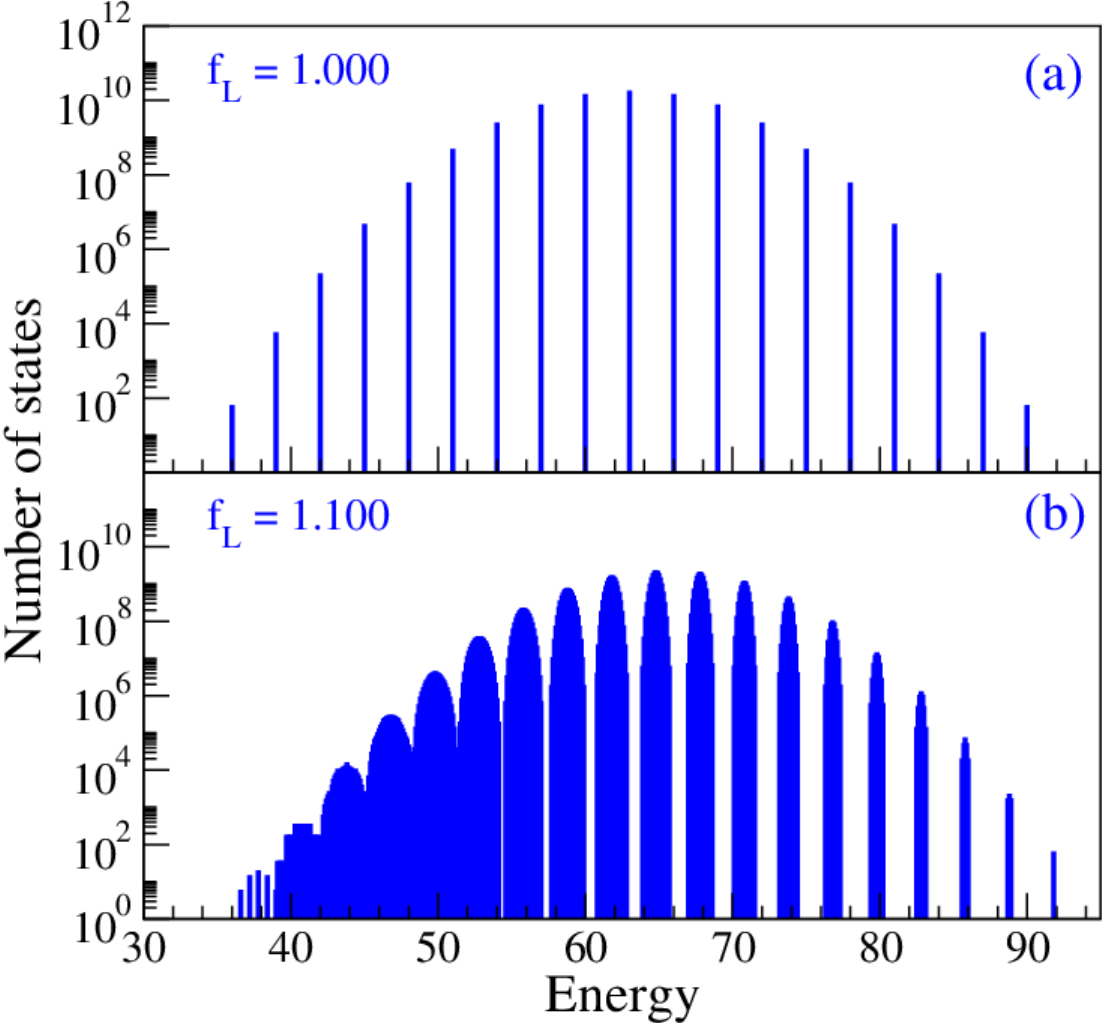}
    \caption{Number of states as a function of energy for $n=m=6$, $g_L=1.000$, and $f_U=g_U=2.000$. Panel (a): $f_L=1.000$; Panel (b): $f_L=1.100$.
 For details, see the text.} 
    \label{fig:dos} 
\end{figure}

This behavior can be understood by examining the number of states of the system at low energies.
Panel (a) of Fig.~\ref{fig:dos} shows the results for $f_L=1$, which corresponds to the case where $\xi^{-1}\rightarrow 0$ as $T\rightarrow 0$.
Here we see that the ground state is degenerated and the energy landscape consists of narrow and well-spaced peaks.
This allows the system to enter an ordered phase with residual entropy at low temperatures, since the Boltzmann factor will not give important contributions above the ground state due to the large barrier.
By contrast, for $f_L=1.1$, the system behaves like a spin glass.
Indeed, panel (b) of this figure shows, in turn, a very complex energy landscape, exhibiting many levels close to each other, almost continuously over a large energy range.
The existence of many states near the ground state, separated by very low barriers, causes the system to become trapped in different states as the temperature is lowered, preventing it from reaching the true ground state.
This leads to a disordered phase, which is a typical characteristic of spin glasses.

\begin{figure}[t] 
    \centering
    \includegraphics[width=0.4825\textwidth]{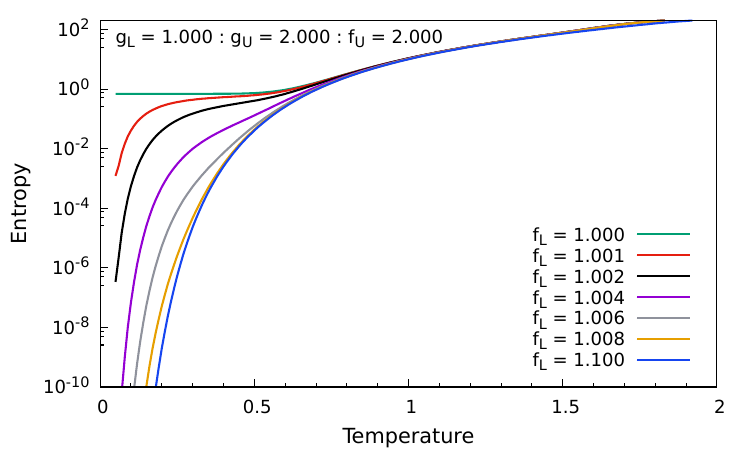}
    \caption{Entropy as a function of temperature for $n=16$.  
 For details, see the text.} 
    \label{fig:Entropy} 
\end{figure}

The above discussion is based on $n=m=6$.
We did not extend these calculations to larger systems due to the huge increase in the number of states with system size.
Reliable estimates, such as those provided by the Wang-Landau method, are beyond the scope of the present work and we simply generated the states one by one.
The use of larger systems would lead to lower energy barriers, which would corroborate our conclusions.
The sensitivity of the energy landscape to $f_L$ may be inferred from Fig.~\ref{fig:Entropy}, which shows the entropy
\begin{equation}
S=-\left(\frac{\partial F}{\partial T}\right)_V
\label{eq:Entropy}
\end{equation}
as a function of temperature for $n=16$ and different values of $f_L$, while keeping the other parameters fixed.
Owing to numerical limitations, the calculations do not include $T=0$ and have been carried out down to the smallest possible $T$ values.
It is clear that the ground state is degenerate for $f_L=1$ and that the departures from the flat behavior of $S$ appear even for very small deviations from $f_L=1$.
The qualitative conclusions obtained with the calculations employing $n=6$ still hold in the present case.

\begin{figure}[t] 
    \centering
    \includegraphics[width=0.4825\textwidth]{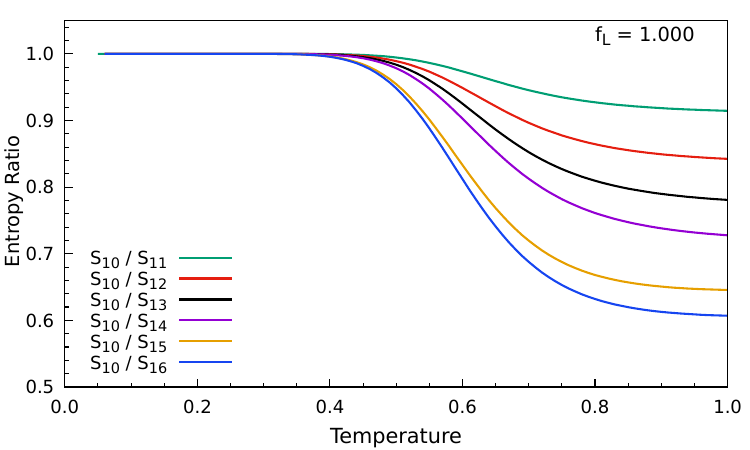}
    \caption{Ratio of the entropy at $n=10$ to the entropy of different system sizes versus temperature. The parameters are set to $f_L=g_L=1.000$, $f_U=g_U=2.000$.
 For details, see the text.} 
    \label{fig:Entropy2} 
\end{figure}

The residual entropy seen in Fig.~\ref{fig:Entropy} for $f_L=1$ is also observed in larger systems.
This is illustrated in Fig.~\ref{fig:Entropy2}, in which we plot the ratio of the entropy for $n=10$ and larger sizes.
The results show that the ratios are equal to 1 up to $T\approx 0.4$, corresponding to the point at which the entropy deviates from the constant value in Fig.~\ref{fig:Entropy} for $f_L=1$.
We therefore conclude that $f_L=1$ acts as a phase boundary separating two qualitatively different behaviors: An ordered phase with residual entropy for $f_L=1$ and a disordered spin glass-like phase for $f_L>1$.

\begin{figure}[h] 
    \centering
    \includegraphics[width=0.4825\textwidth]{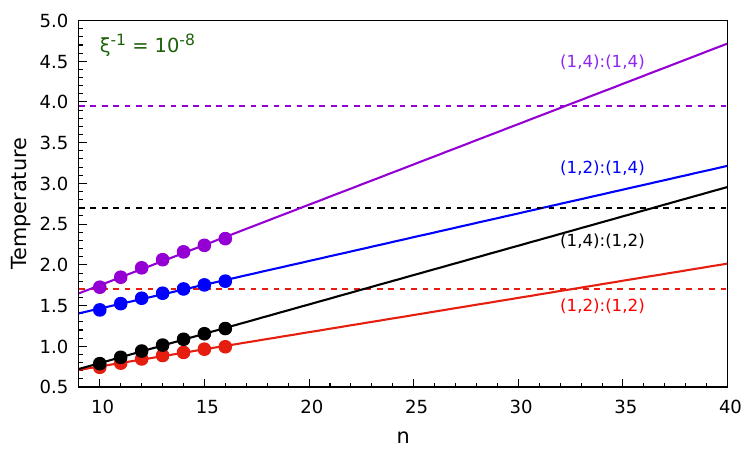}
    \caption{Temperature at which $\xi^ {-1}=10^{-8}$ for different system sizes and parameter sets $(f_L,f_U):(g_L,g_U)$. The horizontal lines correspond to the critical temperatures deduced from the heat capacity. One should note that $T_c$ is the same for $(1,2):(1,4)$ and $(1,4):(1,2)$, and for this reason, the corresponding horizontal lines coincide.
 For details, see the text.} 
    \label{fig:spacing} 
\end{figure}

Finally, Figs.~\ref{fig:cv12_12}--\ref{fig:cv12_14} suggest that the critical temperature extracted from the $\xi^{-1}$ vs.~ $T$ plot would converge to the values obtained in the plots of $C_V$ vs.~ $T$.
We have examined this point by plotting the temperatures at which $\xi^{-1}=10^{-8}$ for different system sizes and parameter sets.
The results are shown in Fig.~\ref{fig:spacing}.
The horizontal lines in this figure correspond to the $T_c$ values obtained from the peaks of $C_V$.
The trends suggest that the curves would intercept the $T_c$ lines, but it is difficult to draw precise conclusions regarding this aspect through the transfer-matrix method due to computational constraints.
Monte Carlo simulations may be helpful in this context but it is beyond the scope of the present study.

\begin{section}{Concluding Remarks}
\label{sect:conclusions}

In this work, we have introduced a new class $C$ of admissible tile types, each
type being a $K$-mer with $1\le K\le\max\{m,n\}-1$, an $m$-cell circular band,
or an $n$-cell circular band. However, $C$ has turned out to
be essentially perfunctory, since what is really at stake is whether all the
different values of $K$ intended for concomitant use can give rise to at least
one tiling for which the property we have called tile maximality holds. While
requiring tile maximality introduces a substantial theoretical innovation into
the study of tilings, pragmatically a central issue related to requiring
maximality has been to demonstrate the use of the transfer-matrix method to
analyze tilings admitting the concomitant use of more than one value of $K$. As
one can see from our comments in Section~\ref{sect:introduction}, thus far this
had not been achieved.

As we have demonstrated in Sections~\ref{sect:theory} and~\ref{sect:results},
the availability of the system's transfer matrix, especially its two largest
eigenvalues, has enabled the identification of a parameter-dependent critical
temperature, at which heat capacity peaks. It has also enabled the use of two
important analytical tools, viz., the system's correlation length and entropy.
For the energy function we adopted, putting all these elements to use has
resulted in the identification of a particular parameterization for which a
phase boundary occurs, separaring an ordered phase from a disordered one.

Moving on to larger systems and possibly different energy functions will likely
require the use of Monte Carlo methods and, with them, the availability of more
powerful computational resources. We believe that our previous experience with
the Wang-Landau method in closely related problems \cite{tiling2025} is bound to
make a difference.

\end{section}

\begin{acknowledgments}
This work was supported in part by the Brazilian agencies Conselho Nacional de Desenvolvimento Cient\'\i­fico e Tecnol\'ogico (CNPq) and Funda\c c\~ao de Amparo \`a Pesquisa do Estado do Rio de Jandiro (FAPERJ), and  the Uruguayan agencies Programa de Desarrollo de las Ciencias B\'asicas (PEDECIBA) and Agencia Nacional de Investigaci\'on e Innovaci\'on (ANII).
This work has been conducted as part of the project INCT-FNA, Proc.~No.~464898/2014-5.

\end{acknowledgments}

\bibliography{tiling}
\bibliographystyle{apsrev4-2}

\end{document}